\RequirePackage{fix-cm}
\documentclass[twocolumn]{svjour3}          % twocolumn
\smartqed  % flush right qed marks, e.g. at end of proof
\usepackage{graphicx}
\journalname{Eur. Phys. J. C}
\usepackage{bm}
\usepackage{slashed}
\usepackage{url}
\usepackage{float}

\usepackage{amsmath}
\usepackage{comment}
\usepackage{subfigure}
\usepackage{float}

\begin{document}

%\preprint{}

\title{Spectrum of a four-dimensional Yang-Mills theory}



\author{Marco Frasca         %\and
        %Second Author %etc.
}


\institute{Marco Frasca \at
              Via Erasmo Gattamelata, 3 \\
							00176 Rome (Italy)
              \email{marcofrasca@mclink.it}           %  \\
}

\date{Received: date / Accepted: date}

\maketitle

\begin{abstract}
We obtain the next-to-leading order correction to the spectrum of a SU(N) Yang-Mills theory in four dimensions and we show agreement well--below 1\% with respect to the lattice computations for the ground state and one of the higher states. 
\end{abstract}

%\pacs{12.38.Aw,11.15.Me}
%\keywords{}

%\section{Introduction}

A successful theoretical approach to treat a Yang-Mills theory in the low-energy limit has been missing so far. It is generally impossible to get the spectrum and the n-point functions unless use is made of lattice computations \cite{Lucini:2004my,Chen:2005mg,Bogolubsky:2007ud,Cucchieri:2007md,Oliveira:2007px}. This difficulty implies that a general understanding from first principles of the behavior of strong interactions is lacking. We are not able to manage QCD in the infrared mostly because perturbation theory fails. The general approach is to use dispersion relations \cite{Narison:2007spa} or phenomenological models obtained from plausibility arguments. These models stay so because of our current knowledge of the 2-point function of the theory relies almost all on lattice computations. We are aware in this way that a mass gap exists, that in the deep infrared the gluon propagator seems to resemble a Yukawa-like one but our inability to do quantum field theory in presence of a large coupling represents a big limitation. 
%Proposals for the gluon propagator in Landau gauge have been appeared in literature \cite{} but these entail to fix several arbitrary parameters so to make them practically useless to obtain predictions to compare with experiment.

In this letter we will show how an approach based on exact solutions of the classical theory, already shown successful for the three-dimensional case \cite{Frasca:2016sky}, can provide agreement with lattice data for the spectrum of a Yang-Mills theory well-below 1\%. Besides, it is possible in this way to perform computations at any order and obtain finite results to compare with experiments without the need to fit a wealth of parameters.

%\section{Gluon propagator \label{sec2}}

For the propagator of Yang-Mills theory in the infrared limit, one generally exploits a current expansion \cite{Frasca:2013tma,Cahill:1985mh}. Instead to start from the action, we prefer the equations of motion \cite{Rubakov:2002fi}
%\begin{strip}
\begin{eqnarray}
&&\partial^\mu\partial_\mu A^a_\nu-\left(1-\frac{1}{\alpha}\right)\partial_\nu(\partial^\mu A^a_\mu)+gf^{abc}A^{b\mu}(\partial_\mu A^c_\nu-\partial_\nu A^c_\mu) \nonumber \\
&&+gf^{abc}\partial^\mu(A^b_\mu A^c_\nu) \nonumber \\ 
&&+g^2f^{abc}f^{cde}A^{b\mu}A^d_\mu A^e_\nu = j^a_\nu.
\end{eqnarray}
%\end{strip}
Then, given a functional form $A^a_\nu=A^a_\nu[j]$ and doing a Taylor expansion around a given asymptotic solution $A^a_\nu[0]$, one has
\begin{eqnarray}
  &&A^a_\nu[j(x)]=A^a_\nu[0]+\int d^dx'\left.\frac{\delta A_\nu^a}{\delta j_\mu^b(x')}\right|_{j=0}j_\mu^b(x') \\
	&&+\frac{1}{2}\int d^dx'd^dx''\left.\frac{\delta^2 A_\nu^a}{\delta j_\mu^b(x')\delta j_\kappa^c(x'')}\right|_{j=0}
	j_\mu^b(x')j_\kappa^c(x'')+\ldots. \nonumber
\end{eqnarray}
Exact solutions can be obtained in this way \cite{Frasca:2015yva} for the Landau gauge. The set of solutions we will start with are
\begin{equation}
    A^a_\nu[0] = \eta_\nu^a\chi(x)
\end{equation}
being $\chi(x)$ a solution of the equation
\begin{equation}
   \partial^2\chi+3Ng^2\chi^3=0
\end{equation}
and this is given by
\begin{equation}
   \chi(x)=\mu(2/Ng^2)^\frac{1}{4}{\rm sn}(p\cdot x+\theta,-1).
\end{equation}
being $\mu$ an arbitrary integration constant having the dimensions of a mass and the ``momenta'' $p$ satisfy the dispersion relation
\begin{equation}
    p^2=\mu^2\sqrt{Ng^2/2}.
\end{equation}
We take these solutions as the ground state of the theory. Then, the propagator of the theory will be
\begin{equation}
    G_{\mu\nu}^{ab}(x,x')=\left.\frac{\delta A_\nu^a(x)}{\delta j_\mu^b(x')}\right|_{j=0}.
\end{equation}
Setting $j=0$ one gets for the Green function of Yang-Mills theory
\begin{eqnarray}
&&\partial^2G_{\nu\rho}^{ae}(x,x')
-\left(1-\frac{1}{\alpha}\right)\partial_\nu\partial^\mu G_{\mu\rho}^{ae}(x,x') \nonumber \\
&&+gf^{abc}G_{\mu\rho}^{be}(x,x')\left(\partial^\mu A^c_\nu-\partial_\nu A^{\mu c}(x)\right) \nonumber \\
&&+gf^{abc}A_\mu^b\left(\partial^\mu G_{\nu\rho}^{ce}(x,x')
-\partial_\nu G_{\mu\rho}^{ce}(x,x')\right) \nonumber \\
&&+gf^{abc}\partial^\mu \left(A^c_\nu G_{\mu\rho}^{be}(x,x')\right) \nonumber \\
&&+gf^{abc}\partial^\mu\left(A_\mu^b G_{\nu\rho}^{ce}(x,x')\right) \\ \nonumber
&&+g^2f^{abc}f^{cdh}G_{\mu\rho}^{be}(x,x')A^{\mu d} A^h_\nu \\ \nonumber
&&+g^2f^{abc}f^{cdh}A^{b\mu}G_{\mu\rho}^{de}(x,x') A^h_\nu \\ \nonumber
&&+g^2f^{abc}f^{cdh}A^{b\mu}A^d_\mu G_{\nu\rho}^{he}(x,x')
= \delta_{ae}\eta_{\nu\rho}\delta^d(x-x').
\end{eqnarray}
or
\begin{eqnarray}
&&\partial^2G_{\nu\rho}^{ae}(x,x')
-\left(1-\frac{1}{\alpha}\right)\partial_\nu\partial^\mu G_{\mu\rho^{ae}}(x,x') \nonumber \\
&&+gf^{abc}G_{\mu\rho}^{be}(x,x')\left(\partial^\mu (\eta^c_\nu\chi(x))-\partial_\nu(\eta^{\mu c}\chi(x))\right) \nonumber \\
&&+gf^{abc}\eta_\mu^b\chi(x)\left(\partial^\mu G_{\nu\rho}^{ce}(x,x')
-\partial_\nu G_{\mu\rho}^{ce}(x,x')\right) \nonumber \\
&&+gf^{abc}\partial^\mu \left(\eta^c_\nu\chi(x) G_{\mu\rho}^{be}(x,x')\right) \nonumber \\
&&+gf^{abc}\partial^\mu\left(\eta_\mu^b\chi(x) G_{\nu\rho}^{ce}(x,x')\right) \\ \nonumber
&&+g^2f^{abc}f^{cdh}G_{\mu\rho}^{be}(x,x')\eta^{\mu d} \eta^h_\nu\chi^2(x) \\ \nonumber
&&+g^2f^{abc}f^{cdh}\eta^{b\mu}G_{\mu\rho}^{de}(x,x') \eta^h_\nu\chi^2(x) \\ \nonumber
&&+g^2f^{abc}f^{cdh}\eta^{b\mu}\eta^d_\mu G_{\nu\rho}^{he}(x,x')\chi^2(x)
= \delta_{ae}\eta_{\nu\rho}\delta^d(x-x').
\end{eqnarray}
We fix the gauge to the Landau gauge ($\alpha=1$) that also grants that we are using exact formulas rather than asymptotic ones. Then,
\begin{equation}
  G_{\mu\nu}^{ab}(x,x')=\delta_{ab}\left(g_{\mu\nu}-\frac{p_\mu p_\nu}{p^2}\right)\Delta(x,x')
\end{equation}
being $p_\mu$ the momentum vector. So, we have to solve the equation
\begin{equation}
   \partial^2\Delta(x,x')+3Ng^2\chi^2(x)\Delta(x,x')=\delta^4(x-x').
\end{equation}
This equation with that of the Green function for the scalar field obtained in \cite{Frasca:2013tma}. Then, the propagator can be immediately written down as \cite{Frasca:2015yva}
\begin{equation}
\label{eq:prop}
   G(p)=\sum_{n=0}^\infty\frac{B_n}{p^2-m_n^2+i\epsilon}
\end{equation}
with
\begin{equation}
    B_n=(2n+1)^2\frac{\pi^3}{4K^3(-1)}\frac{e^{-(n+\frac{1}{2})\pi}}{1+e^{-(2n+1)\pi}}.
\end{equation}
being $K(-1)$ the complete elliptic integral of the first kind and we get the ``mass spectrum''
\begin{equation}
\label{eq:ms}
   m_n=(2n+1)\frac{\pi}{2K(-1)}\mu(Ng^2/2)^\frac{1}{4}.
\end{equation}
This spectrum is kept in quantum field theory but we also obtain higher order corrections. So, our final result for the Green function is
\begin{equation}
    G_{\mu\nu}^{ab}(p)=\delta_{ab}\left(g_{\mu\nu}-\frac{p_\mu p_\nu}{p^2}\right)G(p).
\end{equation}
We see that the Yang-Mills theory shows up a mass gap. We will use the equation for the spectrum to fit with lattice data with a proper quantum correction.

The quantum theory can be studied using Dyson-Schwinger equations for 1- and 2-point functions. These were already discussed in our recent work \cite{Frasca:2015yva}. One has
\begin{eqnarray}
    &&\partial^2G_{1\nu}^{a}(x)+gf^{abc}(
		\partial^\mu G_{2\mu\nu}^{bc}(0)+\partial^\mu G_{1\mu}^{b}(x)G_{1\nu}^{c}(x)- \nonumber \\
		&&\partial_\nu G_{2\mu}^{\nu bc}(0)-\partial_\nu G_{1\mu}^{b}(x)G_{1}^{\mu c}(x) \nonumber \\
		&&+\partial^\mu G_{2\mu\nu}^{bc}(0)+\partial^\mu(G_{1\mu}^{b}(x)G_{1\nu}^{c}(x)))		
		\nonumber \\
		&&+g^2f^{abc}f^{cde}(G_{3\mu\nu}^{\mu bde}(0,0)
		+G_{2\mu\nu}^{bd}(0)G_{1}^{\mu e}(x) \nonumber \\
	&&+G_{2\nu\rho}^{eb}(0)G_{1}^{\rho d}(x)
	+G_{2\mu\nu}^{de}(0)G_{1}^{\mu b}(x)+ \nonumber \\
	&&G_{1}^{\mu b}(x)G_{1\mu}^{d}(x)G_{1\nu}^{e}(x))
		=gf^{abc}(\partial_\nu P^{bc}_2(0)+\partial_\nu (\bar P^{b}_1(x)P^{c}_1(x))) \nonumber \\
	 &&\partial^2 P^{a}_1(x)+gf^{abc}\partial^\mu
	(K^{bc}_{2\mu}(0) \nonumber \\
	&&+P^{b}_1(x)G_{1\mu}^{c}(x))=0.
\end{eqnarray}
The Dyson-Schwinger equations for the two-point functions are
\begin{eqnarray}
\label{eq:ds_3}   
    &&\partial^2G_{2\nu\kappa}^{am}(x-y)+gf^{abc}(
		\partial^\mu G_{3\mu\nu\kappa}^{bcm}(0,x-y) \nonumber \\
		&&+\partial^\mu G_{2\mu\kappa}^{bm}(x-y)G_{1\nu}^{c}(x)
		+\partial^\mu G_{1\mu}^{b}(x)G_{2\nu\kappa}^{cm}(x-y) \nonumber \\
		&&-\partial_\nu G_{3\mu\kappa}^{\mu bcm}(0,x-y)-\partial_\nu G_{2\mu\kappa}^{bm}(x-y)G_{1}^{\mu c}(x) \nonumber \\
		&&-\partial_\nu G_{1\mu}^{b}(x)G_{2\kappa}^{\mu cm}(x-y)
		\nonumber \\
		&&+\partial^\mu G_{3\mu\nu\kappa}^{bcm}(0,x-y)
		+\partial^\mu(G_{2\mu\kappa}^{bm}(x-y)G_{1\nu}^{c}(x)) \nonumber \\
		&&				+\partial^\mu(G_{1\mu}^{b}(x)G_{2\nu\kappa}^{cm}(x-y)))
		\nonumber \\
		&&+g^2f^{abc}f^{cde}(G_{4\mu\nu\kappa}^{\mu bdem}(0,0,x-y)
		+G_{3\mu\nu\kappa}^{bdm}(0,x-y)G_{1}^{\mu e}(x) \nonumber \\ 
		&&+G_{2\mu\nu}^{bd}(0)G_{2\kappa}^{\mu em}(x-y)\nonumber \\
	&&+G_{3\nu\rho\kappa}^{acm}(0,x-y)G_{1}^{\rho b}(x)
	+G_{2\nu\rho}^{eb}(0)G_{2\kappa}^{\rho dm}(x-y) \nonumber \\
	&&+G_{2\nu\rho}^{de}(0)G_{2\kappa}^{\rho bm}(x-y)
	+G_{1}^{\mu b}(x)G_{3\mu\nu\kappa}^{dem}(0,x-y)+ \nonumber \\
	&&G_{2\kappa}^{\mu bm}(x-y)G_{1\mu}^{d}(x)G_{1\nu}^{e}(x)+
	G_{1}^{\mu b}(x)G_{2\mu\kappa}^{dm}(x-y)G_{1\nu}^{e}(x) \nonumber \\
	&&+G_{1}^{\mu b}(x)G_{1\mu}^{d}(x)G_{2\nu\kappa}^{em}(x-y)) \nonumber \\
	&&=gf^{abc}(\partial_\nu K^{bcm}_{3\kappa}(0,x-y)+\partial_\nu (\bar P^{b}_1(x)K^{cm}_{2\kappa}(x-y))) \nonumber \\ 
	&&+\partial_\nu (\bar K^{bm}_{2\kappa}(x-y)P^{c}_1(x))) \nonumber \\
	&&+ \delta_{am}g_{\nu\kappa}\delta^4(x-y) \nonumber \\
	 &&\partial^2 P^{am}_2(x-y)+gf^{abc}\partial^\mu
	(K^{bcm}_{3\mu}(0,x-y) \nonumber \\
	&&+P^{bm}_2(x-y)G_{1\mu}^{c}(x)+ \nonumber \\
	&&P^{b}_1(x)K_{2\mu}^{cm}(x-y))=\delta_{am}\delta^4(x-y) \nonumber \\
	&&\partial^2 K^{am}_{2\kappa}(x-y)+gf^{abc}\partial^\mu
	(L^{bcm}_{2\mu\kappa}(0,x-y) \nonumber \\
	&&+K^{bm}_{2\kappa}(x-y)G_{1\mu}^{c}(x) \nonumber \\
	&&+P^{b}_1(x)G_{2\mu\kappa}^{cm}(x-y))=0. 
\end{eqnarray}

The solutions to this set of equations can be obtained by choosing
\begin{equation}
    G_{1\nu}^{a}(x)=\eta_\nu^a\phi(x)
\end{equation}
being $\eta_\nu^a$ a set of constants and $\phi(x)$ the solution of a differential equation we are going to determine. Besides,  for the Fourier transform of the 2-point function is
\begin{equation}
   G_{2\nu\kappa}^{am}(p)=\delta_{am}\left(g_{\nu\kappa}-\frac{p_\nu p_\kappa}{p^2}\right)\Delta(p)
\end{equation}
with the equation for $\Delta(x-y)$ given below. This set can be solved by taking for the ghost 2-point function
\begin{equation}
    P^{am}_2(p)=\frac{\delta_{am}}{p^2+i0}.
\end{equation}
The ghost field decouples in this case and is free. Then, the 1-point function is obtained by the equation
\begin{eqnarray}
    &&\eta_{\nu}^{a}\partial^2\phi(x)
		+g^2f^{abc}f^{cde}(G_{2\mu\nu}^{bd}(0)\eta^{\mu e}\phi(x) \nonumber \\
	&&+G_{2\nu\rho}^{eb}(0)\eta^{\rho d}\phi(x)
	+G_{2\mu\nu}^{de}(0)\eta^{\mu b}\phi(x) \nonumber \\
	&&+\eta^{\mu b}\eta_{\mu}^{d}\eta_{\nu}^{e}\phi^3(x))=0.
\end{eqnarray}
This becomes the equation for the scalar field, given SU(N) for the gauge group,
\begin{equation}
\label{eq:G1}
   \partial^2\phi(x)+2Ng^2\delta\mu^2\phi(x)+Ng^2\phi^3(x)=0
\end{equation}
having set
\begin{equation}
  \delta\mu^2=\int\frac{d^4p}{(2\pi)^4}\Delta(p),
\end{equation}
again for a mass correction to be evaluated once the propagator is known. 
%For the 2-point function, one has
%\begin{eqnarray}
%    &&\partial^2G_{2\nu\kappa}^{am}(x-y)+gf^{abc}(
%		\partial^\mu G_{3\mu\nu\kappa}^{bcm}(0,x-y)+\partial^\mu G_{2\mu\kappa}^{bm}(x-y)G_{1\nu}^{c}(x)
%		+\partial^\mu G_{1\mu}^{b}(x)G_{2\nu\kappa}^{cm}(x-y) \nonumber \\
%		&&-\partial_\nu G_{3\mu\kappa}^{\mu bcm}(0,x-y)-\partial_\nu G_{2\mu\kappa}^{bm}(x-y)G_{1}^{\mu c}(x) 
%		-\partial_\nu G_{1\mu}^{b}(x)G_{2\kappa}^{\mu cm}(x-y)
%		\nonumber \\
%		&&+\partial^\mu G_{3\mu\nu\kappa}^{bcm}(0,x-y)
%		+\partial^\mu(G_{2\mu\kappa}^{bm}(x-y)G_{1\nu}^{c}(x))
%						+\partial^\mu(G_{1\mu}^{b}(x)G_{1\nu\kappa}^{cm}(x-y)))
%		\nonumber \\
%		&&+g^2f^{abc}f^{cde}(G_{4\mu\nu\kappa}^{\mu bdem}(0,0,x-y)
%		+G_{3\mu\nu\kappa}^{bdm}(0,x-y)G_{1}^{\mu e}(x) 
%		+G_{2\mu\nu}^{bd}(0)G_{2\kappa}^{\mu em}(x-y)\nonumber \\
%	&&+G_{3\nu\rho\kappa}^{acm}(0,x-y)G_{1}^{\rho b}(x)
%	+G_{2\nu\rho}^{eb}(0)G_{2\kappa}^{\rho dm}(x-y)
%	+G_{2\nu\rho}^{de}(0)G_{2\kappa}^{\rho bm}(x-y)
%	+G_{1}^{\mu b}(x)G_{3\mu\nu\kappa}^{dem}(0,x-y)+ \nonumber \\
% &&G_{2\kappa}^{\mu bm}(x-y)G_{1\mu}^{d}(x)G_{1\nu}^{e}(x)+
%	G_{1}^{\mu b}(x)G_{2\mu\kappa}^{dm}(x-y)G_{1\nu}^{e}(x)+
%	G_{1}^{\mu b}(x)G_{1\mu}^{d}(x)G_{2\nu\kappa}^{em}(x-y)) \nonumber \\
%	&&	=gf^{abc}(\partial_\nu K^{bcm}_{3\kappa}(0,x-y)+\partial_\nu (\bar P^{b}_1(x)K^{cm}_{2\kappa}(x-y))) 
%	+\partial_\nu (\bar K^{bm}_{2\kappa}(x-y)P^{c}_1(x)))
%	+ \delta_{am}g_{\nu\kappa}\delta^4(x-y)
%\end{eqnarray}
Then, the equation for the 2-point function becomes \cite{Frasca:2015yva}
\begin{equation}
\label{eq:G2}
   \partial^2\Delta(x-y)+2Ng^2\delta\mu^2\Delta(x-y)+3Ng^2\phi^2(x)\Delta(x-y)=\delta^4(x-y)
\end{equation}
and we see that the mass correction $\delta\mu^2$ is here too. This will provide an equation for the renormalization of mass. Solving eqs.(\ref{eq:G1})-(\ref{eq:G2}), one has
\begin{eqnarray}
    G_1(x)&=&\sqrt{\frac{2\mu^4}{m^2+\sqrt{m^4+2\lambda\mu^4}}}\times \nonumber \\
		&&{\rm sn}\left(p\cdot x+\chi,
		\frac{m^2-\sqrt{m^4+2Ng^2\mu^4}}{m^2+\sqrt{m^4+2Ng^2\mu^4}}\right)
\end{eqnarray}
being $\mu$ and $\chi$ two arbitrary integration constants and we have set $m^2=2Ng^2G_2(0)$, $G_3(0,0)=0$ and taken the momenta $p$ in such a way that
\begin{equation}
\label{eq:disp}
    p^2=m^2+\frac{Ng^2\mu^4}{m^2+\sqrt{m^4+2Ng^2\mu^4}}.
\end{equation}
From these results, we can obtain the correction to the mass spectrum by changing the modulus of the Jacobi elliptic functions and integrals going from $k^2=-1$ to $k^2=\frac{m^2-\sqrt{m^4+2Ng^2\mu^4}}{m^2+\sqrt{m^4+2Ng^2\mu^4}}$ so that, given the dispersion relation in eq.(\ref{eq:disp}), we will get
\begin{equation}
\label{eq:spec}
    m_n(m)=(2n+1)\frac{\pi}{2K(k^2)}\sqrt{m^2+\frac{Ng^2\mu^4}{m^2+\sqrt{m^4+2Ng^2\mu^4}}}
\end{equation}
and, as usual, $\mu$ is an integration constant, having the dimensions of a mass, coming from the integration of the classical theory. This implies that the equation for $\delta\mu^2$ will have the unknown on both sides. So, one can solve it iteratively by taking at the leading order $\delta\mu^2=0$. One has from eq.(\ref{eq:prop})
\begin{eqnarray}
   &&m^2=2Ng^2\int\frac{d^4p}{(2\pi)^4}\sum_{n=0}^\infty(2n+1)^2\frac{\pi^3}{4K^3(-1)}
	\frac{e^{-(n+\frac{1}{2})\pi}}{1+e^{-(2n+1)\pi}}\times \nonumber \\
	&&\frac{1}{p^2-m_n^2(0)+i0}.
\end{eqnarray}
This is the first iterate. In this way, we can evaluate a first correction to the mass spectrum. We just observe that this integral diverges. Indeed, this integral is very well-known in literature and can be evaluated by dimensional regularization. One will get for the finite part
\begin{equation}
    m^2=Ng^2(\gamma-1)\sum_{n=1}^\infty(2n+1)^2\frac{\pi}{32K^3(-1)}\frac{e^{-(n+\frac{1}{2})\pi}}{1+e^{-(2n+1)\pi}}m^2_n(0)
\end{equation}
being $\gamma$ the Euler-Mascheroni constant that we will use to evaluate the full spectrum of the theory. It is interesting to note that this contribution is small and negative. We can write it as
\begin{equation}
   m^2={\bar m}^2Ng^2\sigma
\end{equation}
where we have fixed the string tension to $\sigma=\sqrt{Ng^2/2}\mu^2$ as usual. Then, ${\bar m}^2=-0.03212775693\ldots$. Considering this a small correction to the classical result, the spectrum (\ref{eq:spec}) can be expressed through the formula
\begin{eqnarray}
    &&\frac{m_n}{\sqrt{\sigma}}\approx(2n+1)\frac{\pi}{2K(-1)}\times \nonumber \\
		&&\left[1+\left(\frac{1}{4}-
		\frac{1}{2}\frac{K(\sqrt{2}/2)-E(\sqrt{2}/2)}{K(\sqrt{2}/2)}
		\right){\bar m}^2Ng^2\right].
\end{eqnarray}
Now, we can compare this spectrum with the result from lattice computations given in \cite{Lucini:2004my}. In order to do this, we can fix the value of $Ng^2$ with the value of $\beta=2N/g^2$ used in \cite{Lucini:2004my} to compute the spectrum. So, we rewrite
\begin{eqnarray}
    &&\frac{m_n}{\sqrt{\sigma}}\approx(2n+1)\frac{\pi}{2K(-1)}\times \nonumber \\
		&&\left[1+\left(\frac{1}{4}-
		\frac{1}{2}\frac{K(\sqrt{2}/2)-E(\sqrt{2}/2)}{K(\sqrt{2}/2)}
		\right){\bar m}^2\frac{2N^2}{\beta}\right].
\end{eqnarray}
This yields the comparison table \ref{tab:0++} for the ground state 0$^{++}$ of the theory as seen on lattice computations.
\begin{table}[H]
\begin{center}
\begin{tabular}{|c|c|c|c|c|} \hline\hline
$N$      & Lattice    & Theoretical    & $\beta$     & Error \\ \hline
2        & 3.78(7)    & 3.550927197    & 2.4265      & 6\% \\ \hline 
3        & 3.55(7)    & 3.555252334    & 6.0625      & 0.1\% \\ \hline
4        & 3.56(11)   & 3.556337890    & 11.085      & 0.1\% \\ \hline
6        & 3.25(9)    & 3.557102106    & 25.452      & 8.6\% \\ \hline
8        & 3.55(12)   & 3.557471208    & 45.70       & 0.2\% \\ \hline\hline
\end{tabular}
\caption{\label{tab:0++} Comparison for the ground state at varying $N$. The lattice data are obtained from Ref.~\cite{Lucini:2004my} for the continuum limit.}
\end{center}
\end{table}
The agreement is exceedingly striking for N=3,4 and 8 and well below 1\%. For N=2 and 6 is just a few percent. A similar situation happens for what is labeled as a 2$^{++}$ resonance in \cite{Lucini:2004my}. In this case we get the table \ref{tab:2++}.
\begin{table}[H]
\begin{center}
\begin{tabular}{|c|c|c|c|c|} \hline\hline
$N$      & Lattice    & Theoretical    & $\beta$     & Error \\ \hline
2        & 5.45(11)   & 4.734569596    & 2.4265      & 13\% \\ \hline 
3        & 4.78(9)    & 4.740336446    & 6.0625      & 0.8\% \\ \hline
4        & 4.85(16)   & 4.741783854    & 11.085      & 2\% \\ \hline
6        & 4.73(15)   & 4.742802808    & 25.452      & 0.3\% \\ \hline
8        & 4.73(22)   & 4.743294944    & 45.70       & 0.3\% \\ \hline\hline
\end{tabular}
\caption{\label{tab:2++} Comparison for the 2$^{++}$ state at varying $N$. The lattice data are obtained from Ref.~\cite{Lucini:2004my} for the continuum limit.}
\end{center}
\end{table}
Also in this case the agreement is really stunning. We notice that, in our computations, the dependence on the degree of the group is weak but otherwise noticeable. This is due to the need to perform the comparison exactly for the same $\beta$ as in \cite{Lucini:2004my} for consistency reasons.

%\section{Conclusions \label{sec6}}

We have shown how the spectrum of a Yang-Mills theory in four dimensions can be predicted with an exquisite precision granting strict agreement with lattice computations. This could pave the way for similar precise computations in QCD for the properties of hadrons and, more generally, to get predictions in the spectrum of this theory for exotic states.

%We have shown that, in the framework of our formalism,
% Added on 20 March 2017
%marginal
% 
%confinement is achieved for QCD in three dimensions. We have found extensive agreement with lattice data and preceding theoretical works. We have also shown that numerical factors arbitrarily introduced in some models are completely justified by the set of classical solutions we have chosen to start with. The
% Added on 20 March 2017
%exceptionally good
% 
%agreement between lattice data and theoretical predictions we achieved in the present case can serve as a justification {\sl a posteriori} for the choice of the solutions to start quantum field theory.

%\begin{acknowledgments}
%I would like to thank Marco Ruggieri for useful comments and discussions and for providing code for numerical solving Dyson-Schwinger equations. I also thank Orlando Oliveira for sharing his numerical data with me. This problem was pointed out to me by Owe Philipsen at Bari Conference SM\&FT 2011. I would like to thank him for this.
%\end{acknowledgments}

\end{document}